\documentclass[12pt,preprint]{aastex}
\shorttitle{DoubleClump}
\shortauthors{Nataf et al.}
\usepackage{subfigure}
\usepackage{array,longtable}
\usepackage{natbib}
\usepackage{float}
\usepackage{sidecap}

\begin{document}

\title{The Split Red Clump of the Galactic Bulge from OGLE-III}

\author{D. M. Nataf\altaffilmark{1}, A.Udalski\altaffilmark{2,3}, A.  Gould\altaffilmark{1}, P. Fouqu\'e\altaffilmark{4}, K.~Z. Stanek\altaffilmark{1}}
\altaffiltext{1}{Department of Astronomy, Ohio State University, 140 W. 18th Ave., Columbus, OH 43210}
\altaffiltext{4}{LATT, Universit\'e de Toulouse, CNRS, 14 avenue Edouard Belin, 31400 Toulouse, France}
\altaffiltext{3}{Optical Gravitational Lens Experiment (OGLE)}
\altaffiltext{2}{Warsaw University Observatory, Al. Ujazdowskie 4, 00-478 Warszawa,Poland}
\email{nataf@astronomy.ohio-state.edu, gould@astronomy.ohio-state.edu}

\begin{abstract}
The red clump is found to be split into two components along several sightlines toward the Galactic Bulge. This split is detected with high significance toward the areas $(-3.5<l<1,b<-5)$ and $(l,b)=(0,+5.2)$, i.e., along the Bulge minor axis and at least 5 degrees off the plane. The fainter (hereafter ``main'') component is the one that more closely follows the distance-longitude relation of the Bulge red clump. The main component  is $\sim$0.5 magnitudes fainter than the secondary component and with an overall approximately equal population.  For sightlines further from the plane, the difference in brightness increases, and more stars are found in the secondary component than in the main component. The two components have very nearly equal $(V-I)$ color. 
\end{abstract}


\section{Introduction}
\label{sec:Introduction}
Investigations of the red clump (RC) have had a strong impact on studies of the Galactic Bulge. In their analysis of OGLE-I photometry, \citet*{1994ApJ...429L..73S,1997ApJ...477..163S} found an east-west asymmetry in the dereddened apparent magnitudes of the RC, suggesting a bar oriented at 20-30$^\circ$ relative to the line of sight toward the Galactic center (GC), with the near side in the first Galactic quadrant. \citet{2007MNRAS.378.1064R} obtained similar results using the much larger OGLE-II survey, as did \citet{2005MNRAS.358.1309B} using near-IR photometry along the plane. \citet{2005ApJ...630L.149B} and \citet*{2007A&A...465..825C}
independently found evidence for a long-bar component with a length of $\sim$4.5 kpc and a viewing angle of $\sim$45$^\circ$. \citet{2008A&A...491..781C} argued further that a triaxial spheroid component coexists with the bar, and dominates in the range $|l| \lesssim  10^\circ$. Further evidence of multiple dynamical components are the works of  \citet{2001A&A...379L..44A} and \citet{2005ApJ...621L.105N}, who found evidence that the stellar population toward the range $(|l|\lesssim4,b\lesssim1)$ is distributed differently than that of the main bar. 

The OGLE-III survey has the potential to yield some of the most powerful constraints yet on the major questions: Are there one or two Bulge populations? Where does each dominate? What are their viewing angles and characteristic lengths? With $\sim$96 deg$^{2}$ of observations in $V$ and $I$, the OGLE-III survey is the largest photometric database of the Galactic Bulge that consistently penetrates deep enough to isolate the RC in the coordinate range $(|l|<10,2<|b|<7)$. In the course of fitting for the position of the RC we found that there frequently would appear to be two RCs in a sightline, with the best-fit magnitude being near the mean of that of the two components. We concluded that this effect must be better understood before full modelling of the Bulge can be undertaken. 

In this Letter, we describe the double red clump of the Bulge in the context of our observations: where it is detected, and the relative properties of the two populations in magnitude, color, dispersion and number counts. 

\section{Data}
\label{sec:Data}
\begin{figure}[t]
\includegraphics[totalheight=0.45\textheight]{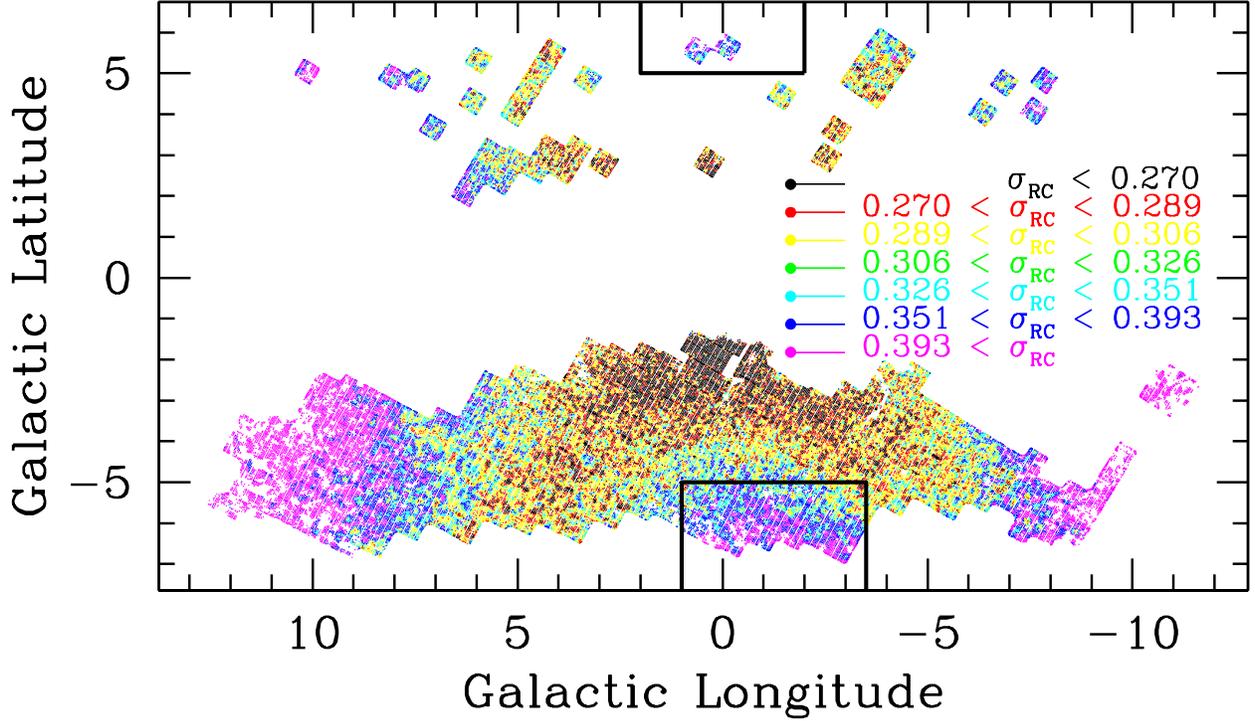}
\caption{The best-fit width of the RC obtained assuming a single Gaussian. The distribution shown as a color-map of septiles of equal areas. The contours of the areas used in this study are delineated by the black rectangles - any OGLE-field with a sub-field centered within those contours is used. Areas where it was difficult to fit for the RC due to factors such as differential reddening are left out as white space. }
\label{Fig:ClumpWidth}
\end{figure}

OGLE-III observations were obtained from the 1.3 meter Warsaw Telescope, located at the Las Campanas Observatory in Chile. The camera has eight 2048x4096 detectors, with a combined field of view of $0.6^{\circ}\times
0.6^{\circ}$ yielding a scale of approximately 0.26$\arcsec$/pixel. We use observations from a mosaic of 267 fields directed toward the Galactic Bulge, which are almost entirely within the range $-10<l<10$ and $2<|b|<7$. The photometric coverage is shown in Figure \ref{Fig:ClumpWidth}. Of the 267 fields, 37 are at northern latitudes. More detailed descriptions of the instrumentation, photometric reductions and astrometric calibrations are available in \citet{2003AcA....53..291U} and \citet{2008AcA....58...69U}.

\section{Method}
We fit for the position of the RC in a multi-step process. We first estimate the position of the RC by eye, by looking at sample locations within every 
OGLE-III field. This is used as a guess for the following step. At each of $\sim$92,000 windows spaced approximately 2$\arcmin$ apart, a geometric centroid in $(I,V-I)$ is calculated based on stars that are within  0.5 magnitudes of the guessed clump location on the color magnitude diagram (CMD). Each location is inspected by eye to verify the fit, and the process is repeated until every inspection window is found to have a visually acceptable RC centroid. It was in this step, with the large number of direct visual inspections required, that we first noticed that a number of sightlines had multiple RCs in their CMDs. 

A more detailed analysis is then required, to delineate the peak of the RC population from the peak of the combined RC+RG (Red Giant) population, and to estimate properties such as the number density of RC stars and their dispersion in magnitude within a given sightline. To account for these factors, we select stars for fitting that satisfy the following color-magnitude cut:
\begin{eqnarray}
 -0.2 < (V-I) - (V-I)_{RC,G} <0.2 \nonumber \\
 -1.5 < I - I_{RC,G}  < 1.0
\label{EQ:constraint2}
\end{eqnarray}
where the subscript (RC,G) refers to the guessed RC location inferred from the previous process of geometric centroiding. The RC+RG population is then fit with the equation:
\begin{equation}
N(m) = A\exp\biggl[B(m-m_{RC})\biggl] + \frac{N_{RC}}{\sqrt{2\pi}\sigma_{RC}}\exp \biggl[{-\frac{(m-m_{RC})^2}{2\sigma_{RC}^2}}\biggl],
\label{EQ:Exponential}
\end{equation}
where $N_{RC}$ is the number of RC stars within a window, $m_{RC}$  is their estimated brightness peak, ${\sigma}_{RC}$ is their dispersion in brightness, and $A$ and $B$ fit an exponential profile to the magnitude distribution of RG stars, corresponding to a power-law fit to the flux. The best-fit is obtained by minimizing ${\chi}^2$ over 50 histogram bins in the magnitude range of Equation (\ref{EQ:constraint2}).
\begin{equation}
  {\chi}^2 =  \sum_{i=1}^{50}\frac{{(N_{model}-N_{obs})}^2}{N_{model}}	
\end{equation}
There is a degeneracy that manifests itself as the width of the RC ${\sigma}_{RC}$ gets larger. The distribution of stars can then be well-fit by a range of parameter values involving larger ${\sigma}_{RC}$, larger $N_{RC}$, and lower $B$, i.e., it becomes harder to distinguish between the RC and the RG branch as the RC gets wider. As such, we impose a Gaussian prior, $(\mu,\sigma)_{f_{i}} = (0.3652,0.0269)$,  on the fraction $f_{i}$ of stars in a given sightline that are RC stars divided by the sum of $N_{RC}$ and the RG stars that satisfy the criteria of Equation (\ref{EQ:constraint2}).

\begin{figure}[ht]
\includegraphics[totalheight=0.74\textheight]{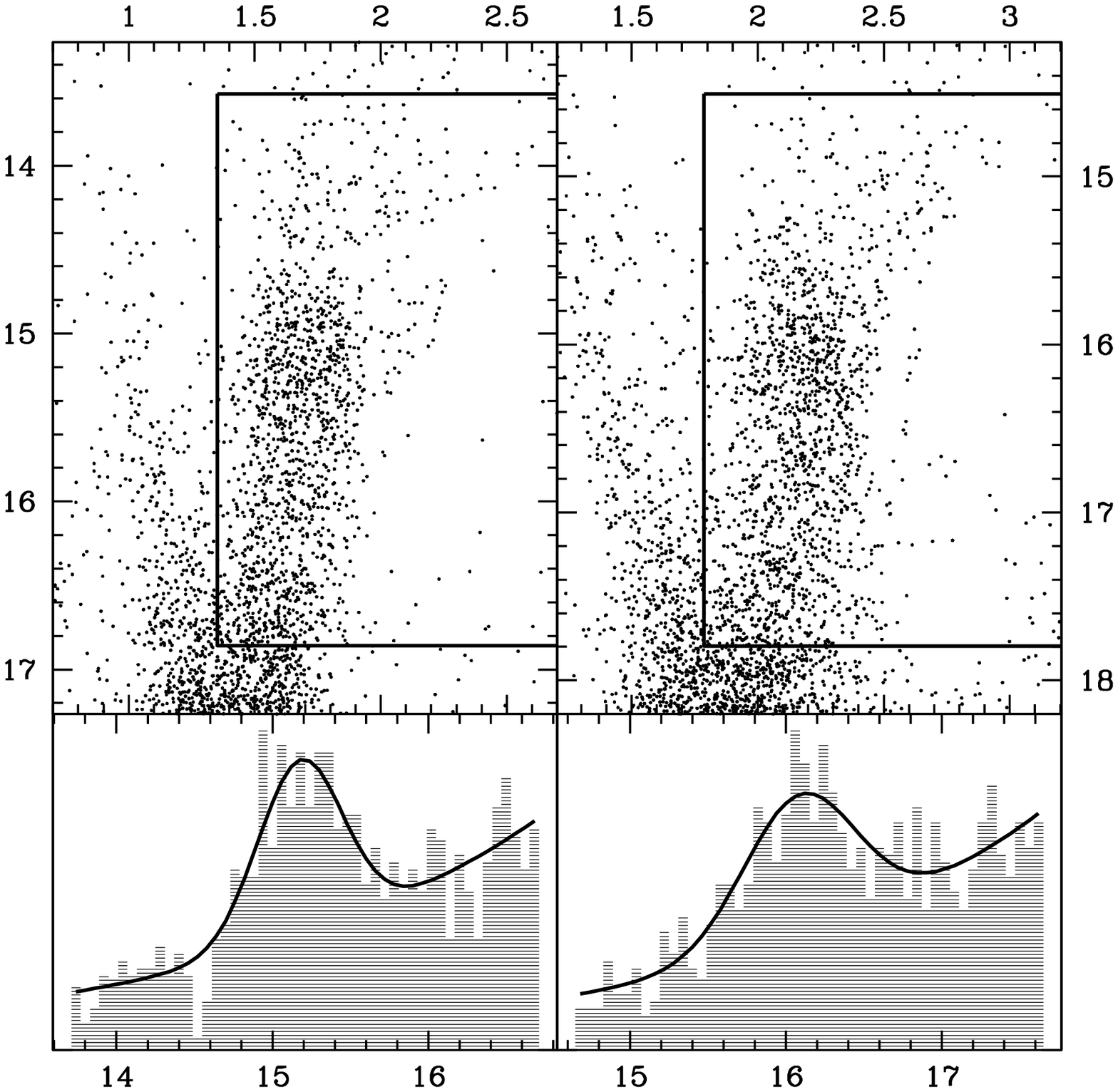}
\caption{Color-Magnitude diagrams and brightness distribution with best-fit single clump model for two locations. Left: $(l,b)=(1.0,-3.9)$, with paramaters $(m_{RC},{\sigma}_{RC},N_{RC},)=(15.177, 0.273, 288.6 )$. Fitting a second Gaussian improves ${\chi}^2$ from 54.3 to 50.7, i.e. 3 additional d.o.f. yields ${\Delta}{\chi}^2=3.6$. Right: $(l,b)=(-4.0,-3.3)$, with clump paramaters $(m_{RC},{\sigma}_{RC},N_{RC},)=(16.085,0.360,310.1)$. ${\Delta}{\chi}^2=3.1$ for a second Gaussian. }
\label{Fig:SingleClump}
\end{figure}

This procedure proves adequate for large swaths of the Galactic Bulge, examples are shown in Figure \ref{Fig:SingleClump}. However it is not adequate for regions where the RC is split into two. 

\subsection{The Split Red Clump}

\begin{figure}[ht]
\includegraphics[totalheight=0.74\textheight]{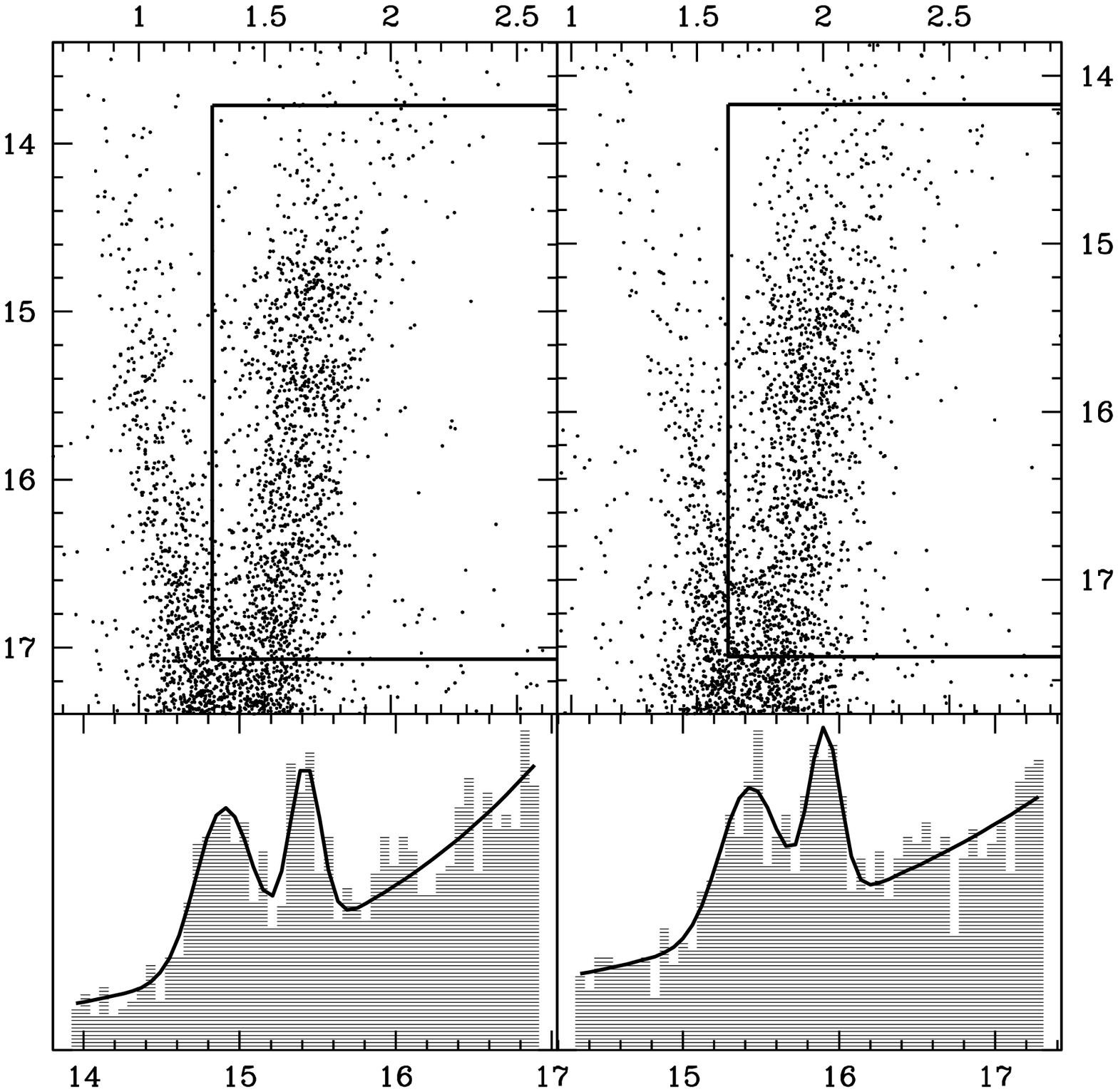}
\caption{Color-Magnitude diagrams and brightness distribution with best-fit double clump model for two locations. Left: $(l,b)=(0.27,-5.77)$, the $\Delta{\chi}^2=26.7$ when a second Gaussian is fit, the parameters are: $(m_{RC,1},{\sigma}_{RC,1},N_{RC,1}, m_{RC,2},{\sigma}_{RC,2},N_{RC,2}) = (14.90, 0.18,166.42, 15.42, 0.10, 98.51,)$. Right:  $(l,b)=(-0.28,5.76)$, the $\Delta{\chi}^2=14.4$. The parameters are: $(m_{RC,1},{\sigma}_{RC,1},N_{RC,1}, m_{RC,2},{\sigma}_{RC,2},N_{RC,2}) = (15.42, 0.19, 147.46,  15.91, 0.10, 94.23,)$.}
\label{Fig:DoubleClump}
\end{figure}

Our first computational evidence of the existence of a split RC is the distribution of ${\sigma}_{RC}$ as a function of longitude and latitude, plotted in Figure \ref{Fig:ClumpWidth}. There is an east-west asymmetry in the southern Galactic hemisphere. In the coordinate range $(1<l<5,b<-5)$, ${\sigma}_{RC}$ is typically $\sim$0.3 magnitudes, whereas for the coordinate range $(-3<l<1,b<-5)$, ${\sigma}_{RC}$ exceeds 0.4 magnitudes. On closer inspection, we observed is that there are two ``bumps'' toward many of these sightlines. The Gaussian fit is then spurious, with its ``peak'' falling in between the two ``real'' peaks, and its width broadened to partially encompass both bumps. Examples are shown in Figure \ref{Fig:DoubleClump}. This effect may also be present in other parts of the sky, but if so the two clumps are likely separated by less than 0.4 magnitudes, rendering it more difficult to distinguish them. 

We decided to focus on those OGLE-III fields where the double RC would be most prominent: BLG125, 126, 127, 128, 134, 135, 136, 137, 143, 144, 145, 146, 151, 152, 153, 154, 159, 160, 161, 162, 167, 168, 169, 170, 176, 177, 178, 186, 355, 357; approximately corresponding to the areas  $(-3.5<l<1,b<-5.0)$ and $(-2<l<2,5<b)$. A full list of OGLE-III Bulge fields and their coordinates can be found on the OGLE webpage\footnote{http://ogle.astrouw.edu.pl/$\sim$ogle/}. 

In each window, we fit the RC+RG population to the following equation:
\begin{equation}
N(m) = A\exp\biggl[B(m-m_{RC,1})\biggl] + \frac{N_{RC,1}}{\sqrt{2\pi}\sigma_{RC,1}}\exp\biggl[{-\frac{(m-m_{RC,1})^2}{2\sigma_{RC,1}^2}}\biggl] + \frac{N_{RC,2}}{\sqrt{2\pi}\sigma_{RC,2}}\exp\biggl[{-\frac{(m-m_{RC,2})^2}{2\sigma_{RC,2}^2}}\biggl]
\label{EQ:Exponential2}
\end{equation}
Within the color-magnitude selection box:
\begin{eqnarray}
 -0.2 < (V-I) - (V-I)_{RC,G}  \nonumber \\
 -1.5 < I - I_{RC,G}  < 1.5
\label{EQ:constraint3}
\end{eqnarray}
The angular radius used is the smallest one such that $\sim$1200 stars meet the conditions of Equation (\ref{EQ:constraint3}), and the coordinates selected for inspection are placed 3$\arcmin$ apart within every OGLE-III subfield - they will be closer if they are at the edges of neighboring fields. Degeneracies and correlations remained a concern in early trial runs, and thus we imposed three priors on the fit. The fraction of stars $f_{i}$ that are RC stars (of either clump) and meet the conditions of Equation (\ref{EQ:constraint3}) was given the Gaussian prior of $(\mu, \sigma)_{f_{i}} = (0.25, 0.05)$; Gaussian priors were also used to discourage solutions with a clump dispersion ${\sigma}_{RC}<0.1$ mag, or a brightness peak either at least 1 magnitude brighter or 0.5 magnitudes fainter than the peak obtained from a single Gaussian fit.


\subsection{Comparing $V-I$ color}

\begin{figure}[ht]
\includegraphics[totalheight=0.74\textheight]{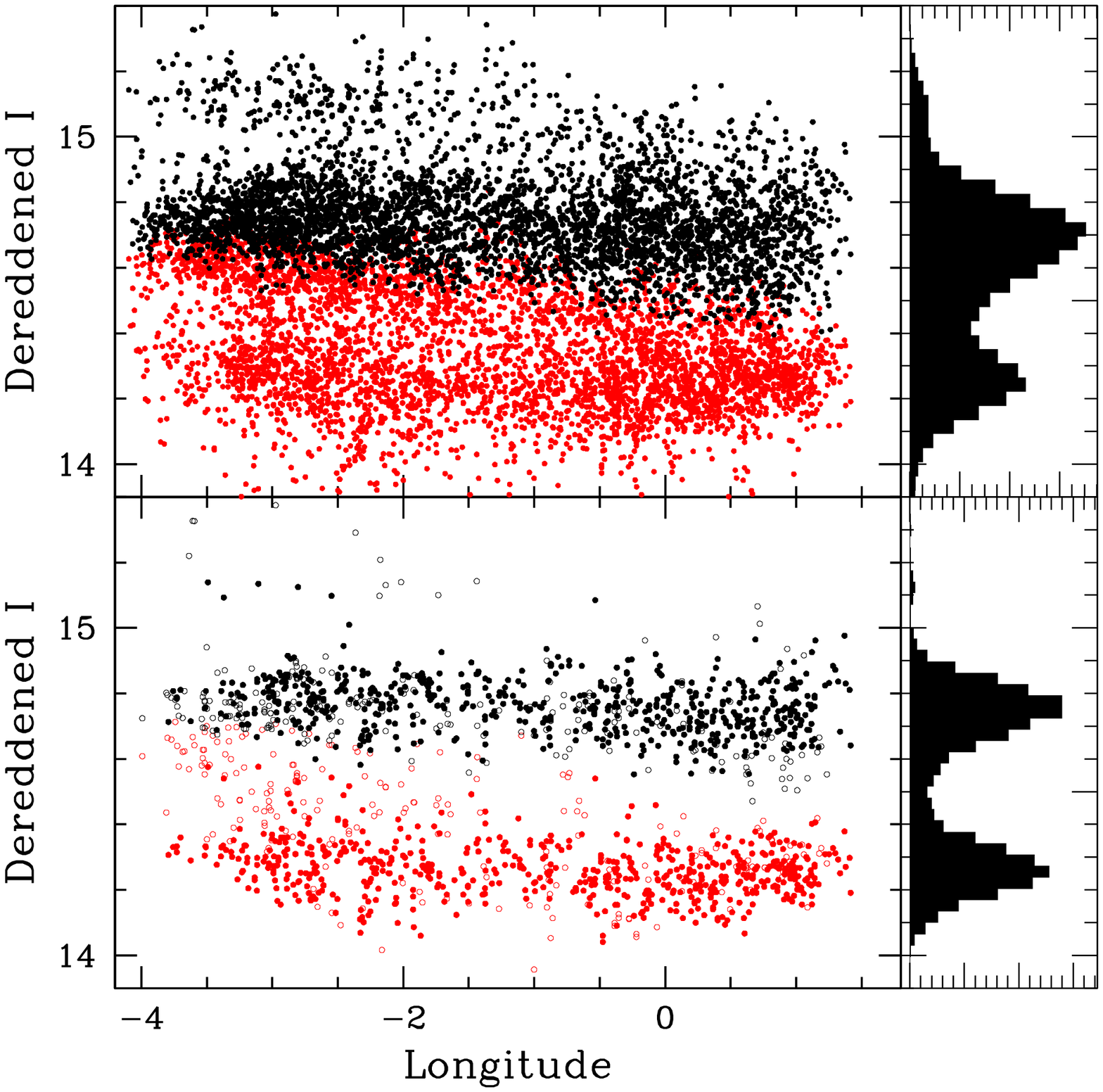}
\caption{TOP: The dereddened main and secondary RC are respectively plotted in black and red as a function of longitude for every sightline. Since a double-RC is not prominent in every sightline a few sequences are seen, including a possible red giant branch bump at $I\sim15.2$. BOTTOM: The plotted points are those for which a second Gaussian yields a $\Delta{\chi}^2 > 8$ and the smaller RC with at least one quarter the population of the larger RC. Those for which the difference in magnitude is $0.4<{\Delta}I<0.6$ are plotted as filled circles. }
\label{Fig:Populations}
\end{figure}

We compare the $V-I$ color of the two RC by inspecting 621 high-confidence sightlines for which $\Delta{\chi}^2 > 8$ when a second Gaussian is added to the fit, the difference in magnitude is $0.4<{\Delta}I<0.6$, and the smaller RC has at least one quarter the population of the larger RC. The effect of these selection criteria is shown in Figure \ref{Fig:Populations}. In each sightline, a box is set with an area of $(0.3 \times0.3)$ mag$^{2}$ in color-magnitude space, centered on each of the two RCs in the window. The difference in color, $(V-I)_{bright}-(V-I)_{faint}$, had a mean of 0.0034 magnitudes and a dispersion of 0.012 magnitudes. It is thus consistent with zero. Values ranged between $-0.028$ mag and $+0.036$ mag, which is roughly the 3$\sigma$ error range, where $\sigma$ is the error in the mean determined determined from the typical dispersion in $V-I$ color of the stars in a box ($\sim$0.08 mag) and the typical number of stars in a box ($\sim$120). 

\subsection{Properties of the Two Red Clumps as a Function of Reddening, Longitude and Latitude}
A least-squares fit is computed for the magnitude as a function of reddening, longitude and absolute latitude on the same 621 selected sightlines, using the following formalism:
\begin{eqnarray}
 I = I_{0} + R_{I}*E(V-I) + \frac{d{\mu}}{dl}*l + \frac{d{\mu}}{d|b|}*(|b|-5),
\label{EQ:VMILatFit}
\end{eqnarray}
where $R_{I}=A_{I}/E(V-I)$ is ratio of total to selective extinction, $E(V-I)$ assumes $(V-I)_{0}=1.08$ \citep{2010A&A...512A..41B}, and $\frac{d{\mu}}{dl}$ and $\frac{d{\mu}}{d|b|}$ parametrize any change in distance modulus to the source populations due to their spatial distribution. The values obtained are: 
\begin{eqnarray}
 I_{main} = 14.735 + 1.168*E(V-I) -0.0136*l + 0.018*(|b|-5) \nonumber \\
I_{secondary} = 14.259 + 1.168*E(V-I) -0.0136*l -0.010*(|b|-5)
\label{EQ:VMILatFit2}
\end{eqnarray}
We fixed $R_{I}$ and $\frac{d{\mu}}{dl}$ to be the same for both populations as they only differed at the $\sim$2.0$\sigma$ and $\sim$1.0$\sigma$ level respectively. By contrast, the latitude terms, differ by $\sim$6.0$\sigma$. The $\sim$2$\sigma$ detection of a difference in $R_{I}$ may be due to systematic effects such as the presence of dust between the two populations.

\section{Results}

We observe the following:
\begin{itemize}
 \item The main clump better traces the dereddened Bulge RC magnitude known from previous studies \citep{1997ApJ...477..163S,2007MNRAS.378.1064R}, though it is $\sim$0.1 magnitudes fainter. 
 \item The secondary RC is about 0.5 magnitudes brighter than the main RC. The difference in brightness increases further from the plane.
 \item The two RC are approximately equally populated, with the secondary RC becoming more and more populated relative to the main RC further from the plane. 
 \item The two RCs have approximately equal $(V-I)$ color.
 \item The main RC becomes narrower for sightlines further from the plane, whereas the secondary RC becomes broader. 
 \item The same trends are observed in the northern hemisphere, but they are less statistically significant since there are 15x fewer data.
 \item The secondary RC becomes more and more populated relative to the main RC as longitude increases.
 \item The difference in brightness $I_{main}-I_{secondary}$, and the ratio of the widths ${\sigma}_{RC,main}:{\sigma}_{RC,secondary}$, show no statistically significant trends with longitude.
 \item Within the OGLE-III Bulge database, the effect is visible in the coordinate range $(-3.5\lesssim l \lesssim1,5 \lesssim |b|)$. If it is present elsewhere, it is less pronounced.
\end{itemize}

\begin{figure}[ht]
\includegraphics[totalheight=0.74\textheight]{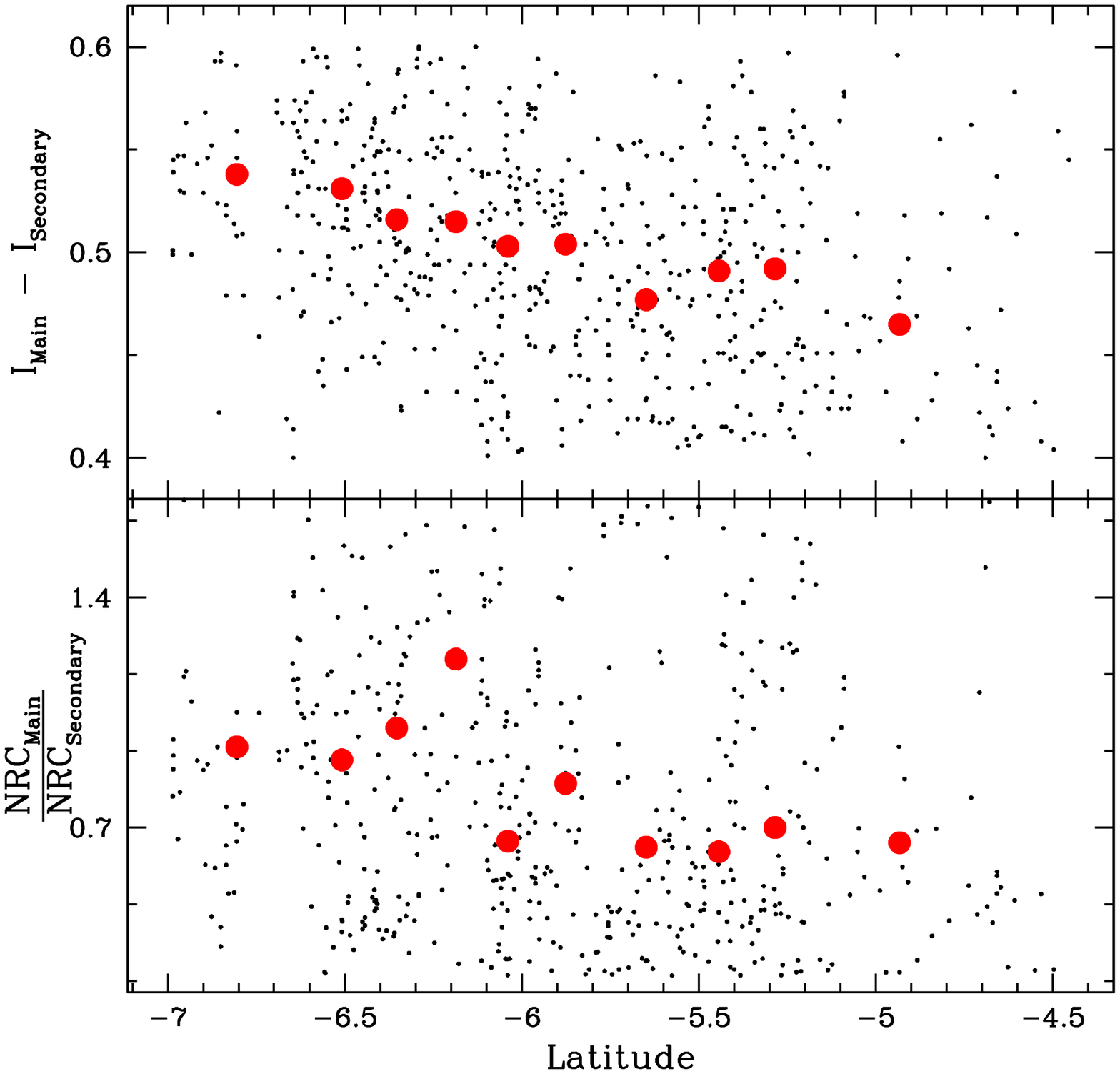}
\caption{The magnitude difference between the bright and faint clump, and the ratio of the two populations as a function of latitude. We plotted points for those 621 high-confidence sightlines described previously. The medians are shown as red points.}
\label{Fig:Geography}
\end{figure}

In Figure \ref{Fig:Geography}, we plot the distribution of the ratio of the populations of the two RCs, $NRC_{main}/NRC_{secondary}$, and the difference in brightness $I_{main}-I_{secondary}$ as functions of southern Galactic Latitude .

\section{Discussion and Conclusion}
\label{sec:Discussion}

It is not clear what this double clump feature could be. \citet{1998mcdg.proc..249G} summarized many of the pit-stops on the red giant branch, such as the RGB-bump and the AGB-bump. The AGB-bump for example, would seem to fit the bill as it is suggested to be of a similar color to the red clump and about 1 magnitude brighter. However, why would it be visible in these specific parts of the sky? It is also not expected to be equally populated to the RC.

An investigation of the RC across the entire OGLE-III Galactic Bulge data set is in progress (Nataf et al. 2010, in preparation). It may be found that the expected brightness at the locations used for this study is in between those of the two RCs. If that is the case, one possible explanation is that one RC belongs to the Galactic bar and the other to the spheroid component of the Bulge, and that due to geometrical effects their separation is more obvious for $(-4<l<1,|b|>5)$. 

It is not even certain that this double RC effect is localized - it is only certain that its \textit{degree of detectability} is localized. We have deviations between the two clumps of between 0.4 and 0.6 magnitudes. If elsewhere on the sky they were separated by 0.4 magnitudes or less, they might be indistinguishable from a single broad clump. Philosophically, the signal is not from two peaks but from a trough between the two peaks. This trough is suppressed by statistical noise and by the underlying RG branch, which is a monotonically increasing function of magnitude. The intermediary distribution between two peaks with a trough and a single peak, i.e., that of a flat distribution over a range of $\sim$0.4 magnitudes, could also be produced by a boxy Bulge. As such, if it were seen it could not be argued to be necessarily due to the presence of the two RCs moving closer together.

The color constraint is a powerful one. \citet{2001MNRAS.323..109G} investigated the behavior of the RC as a function of age and metallicity using the \citet{2000A&AS..141..371G} evolutionary tracks and isochrones. They found that for metallicities ranging from Z = 0.0004 through to Z = 0.03, the $(V-I)$ color levels off after about 3 Gyr. The brighter RC could, for example, be both of a higher metallicity than the regular RC and younger than 2-3 Gyr - the two effects could then cancel out and the difference in $I$-band would be consistent with the observed value of 0.5 magnitudes. If this secondary RC is older than $\sim$3 Gyr, the two populations would need approximately equal metallicity, and the secondary RC would need to be closer to explain the magnitude difference. For example, for respective ages of the populations of 3 and 10 Gyr, the secondary RC would need to be about 1.0 kpc closer. For ages of 7 and 10 Gyr, it would need to be about 1.5 kpc closer. We note that the model results in \citet{2001MNRAS.323..109G} assume a scaled-solar metallicity -- a broader range of choices may be available if [$\alpha$/Fe] is allowed to differ between the two RCs.

A captured and cannibalized dwarf galaxy is another possibility. Since its stars would occupy a specific set of orbits, this could explain why this feature is seen in some parts of the sky and not others. The current nearest galaxy to the MW is the Sagittarius stream \citep{1995MNRAS.277..781I}; it is at an earlier stage of digestion than this hypothetical stream would be. Radial velocities (RV) would likely be significantly different between the two populations in such a scenario and as such RV measurements would provide an effective constraint on the nature of this population. Table \ref{table:surveys} lists 5,085 RC stars including their OGLE ID, astrometric coordinate, photometry and whether or not they are of the main or secondary component. Proper motions (PM) are an alternative option to the same ends. PM measurements are not as precise as those done with RV. The proper motion catalog for OGLE-II \citep{2004MNRAS.348.1439S} includes two fields that overlap with the sightlines used in this study, BUL\_SC6 and BUL\_SC7. A proper reduction catalog of OGLE-III (Poleski et al. 2011, in preparation) will yield measurements for tens of thousands of RC stars in the regions of interest. Currently, we find only one mention of the Bulge double RC in the literature: \citet{2010IAUS..265..271Z} briefly reports on a double RC finding along the bulge minor axis and at distances in excess of $\sim$700 pc removed from the plane. An early analysis of Spitzer-GLIMPSE mid-IR photometry closer to and in the plane \citep{2003PASP..115..953B} also shows the presence of some multiple peaks (Benjamin et al. 2010, in preparation).

We bring up two other ``double RC'' findings in the literature for comparison. The first being the globular cluster Terzan 5 \citep{2009Natur.462..483F} and the other in the Sagittarius stream \citep{2010arXiv1007.3510C}. Neither case is directly analogous to our own. The two RCs in Terzan 5 are separated not just in magnitude, but in color as well, with the fainter RC being about $\sim$0.15 magnitudes bluer in $(J-K)$. This lends itself to an explanation of multiple stellar populations with different ages and iron content, or with similar ages and a different iron as well as helium content \citep{2010ApJ...715L..63D}. The Sagittarius dwarf galaxy on the other hand has a more natural geometric explanation to its double RCs - sightlines intersecting with multiple wraps of stars. \citet{2010arXiv1007.3510C} detect multiple double RCs often separated by 0.5 magnitudes or greater. 

Moving forward, one way to ascertain if this double clump is widespread is to obtain multi-color photometry as well as spectra in regions where the two clumps are distinct. \citet{2002MNRAS.337..332S}, analyzing the results of their simulations in $K$-band, found that the brightness of the clump increases with increasing metallicity, whereas $V$ and $I$-band brightness decrease with increasing metallicity. If distinctions such as an enhanced metallicity appear, they could be used to confirm or rule out the possibility that this double RC is more common in the Bulge than just in the regions of interest where we verified its presence. The implications would be widespread, as the width of the RC, ${\sigma}_{RC}$ is used to constrain models of the Bulge morphology \citep{1997ApJ...477..163S,2007MNRAS.378.1064R,2008A&A...491..781C}. If the width is due to the presence of multiple stellar populations, rather than a dispersion due to a range of distances, then the thickness of the Bulge obtained from RC studies would be seriously overestimated and the viewing angle of the bar would be underestimated. Of equal importance, variations in apparent magnitude would then often be due to the varying relative size of the two populations, rather than the a change in distance modulus due to the orientation of the bar. 

\acknowledgments
DMN and AG were partially supported by the NSF grant AST-0757888. The OGLE project has received funding from the European Research Council
under the European Community's Seventh Framework Programme
(FP7/2007-2013) / ERC grant agreement no. 246678 to AU.


\begin{table*}[h]
\caption{5085 RC stars have been tabulated from 21 high-confidence sightlines to facilitate further study. A briefer version of the table is shown below. Selected stars had a $(V-I)$ within 0.15 magnitudes that of the fit, and were assigned ``Main'' or ``Secondary'' status if their I-band brightness was within 0.15 magnitudes of the fit value. The full table is available at: ftp://ftp.astronomy.ohio-state.edu/pub/nataf/DoubleClumpList.txt}
\begin{center}
\scalebox{0.95}
{\begin{tabular}{ l l l l l l l}
\\
\hline\hline
Subfield & OGLE ID & STATUS & RA & DEC & V-I & I \\
\hline
128.1 & 26929 & Main & 18:06:10.60 & -35:02:12.2 & 1.293 & 15.195 \\
128.1 & 26931 & Secondary & 18:06:11.71 & -35:02:29.0 & 1.426 & 14.754 \\
128.1 & 26932 & Secondary & 18:06:11.82 & -35:02:26.7 & 1.323 & 14.693 \\
128.1 & 26933 & Secondary & 18:06:12.34 & -35:03:40.2 & 1.332 & 14.774 \\
128.1 & 26934 & Main & 18:06:12.43 & -35:03:33.0 & 1.519 & 15.261 \\
128.1 & 26936 & Main & 18:06:13.65 & -35:01:39.5 & 1.375 & 15.222 \\
128.1 & 26938 & Main & 18:06:13.96 & -35:03:04.7 & 1.504 & 15.259 \\
128.1 & 26940 & Secondary & 18:06:15.25 & -35:02:10.9 & 1.375 & 14.645 \\
128.1 & 26944 & Main & 18:06:15.77 & -35:02:39.1 & 1.469 & 15.327 \\
\hline
\end{tabular}
\label{table:surveys}
}\end{center}
\end{table*}

\end{document}